\documentclass[review]{elsarticle}
\usepackage{amsmath}
\usepackage{amsfonts}
\usepackage[german, english]{babel}
\usepackage{graphicx}
\usepackage{amssymb}
\newtheorem{thm}{Theorem}
\newtheorem{lem}[thm]{Lemma}
\newdefinition{rmk}{ }
\newproof{pf}{Proof}
\begin{document}
\begin{frontmatter}
\title{Small perturbations leading to large asymmetries in the probability density of the ground state}
\author{R. Mu\~{n}oz-Vega\corref{cor1}}
\ead{rodrigo.munoz@uacm.edu.mx}
\address{Universidad Aut\'{o}noma de la Ciudad de M\'{e}xico, Fray Servando Teresa de Mier 92, Col Centro, Del Cuauht\'{e}moc, Ciudad de M\'{e}xico, CP 06080, M\'{e}xico}
\author{G. Fern\'{a}ndez-Anaya}
\ead{guillermo.fernandez@ibero.mx}
\address{Universidad Iberoamericana, Prolongaci\'{o}n Paseo de la Reforma 880, Col Lomas de Santa Fe, Del A Obreg\'{o}n, Ciudad de M\'{e}xico, CP 01219, M\'{e}xico} 
\cortext[cor1]{Corresponding author}
\begin{abstract}
It is shown that in one-particle Schr\"{o}dinger quantum mechanics a small perturbation of a one-dimensional potential can produce a large change in the ground state, the effect becoming more pronounced with growing typical length of the potential, contrary to what standard stationary perturbation theory could make one believe.The possible consequences of this result for solid state physics and other fields are also discussed, and examples of such non-standard perturbations are furnished. 
\end{abstract}
\end{frontmatter}
\section{Introduction}
The sole purpose of the Letter at hand is to call attention to following fact, and its consequences: that in a wide family of one-particle one-dimensional models, the ground state can be modified in a dramatic way with only a negligible  perturbation in the potential. Stated in more precise terms, in the following pages it will be shown that for any $\epsilon>0$ and any $M>0$ there exists a family $\{V_{\epsilon,M,\alpha}\}\vert_{\alpha\in(0,1)}$ of analytic potentials, with analytic square-integrable ground eigensolutions $\psi^{(\epsilon,M,\alpha)}$, such that
 \begin{equation} \label{I.02}
 \vert V_{\epsilon,M,\alpha}(x)- V_{\epsilon,M,\beta}(x) \vert< \epsilon \forall x\in\mathbb{R}
 \end{equation}
for all $\alpha,\beta\in(0,1)$ and that there exists  $\beta_{M},\alpha_{M}\in(0,1)$ such that
 \begin{equation} \label{I.03}
\frac{Q[\psi^{(\epsilon,M,\alpha_{M})}]}{Q[\psi^{(\epsilon,M,\beta_{M})}]}>M \quad ,
 \end{equation}
with quotient $Q$ defined as
 \begin{equation} \label{I.04}
Q[f]=\frac{\int_{0}^{\infty}[f(x)]^{2}dx}{\int_{-\infty}^{0}[f(x)]^{2}dx}
 \end{equation}
for any square-integrable $f:\mathbb{R}\rightarrow\mathbb{R}$. 

Stated in more physical terms, the content of the Letter reduces to the following observation: when the stationary ground state $\psi$ of a system can be seen to be the coherent superposition of two states:
\begin{equation} \label{I.05}
\psi=n_{R}\psi^{(R)}+n_{L}\psi^{(L)}\quad \textrm{ , }
\end{equation}
the coefficients $n_{L},n_{R}$ can be changed almost at whim, giving rise to a large asymmetry in the probability distribution, \emph{if} the overlap between this two states is small, and a particular instance of this occurs \emph{when} the distance between the expected positions $\langle\psi^{(R)}\vert x\vert\psi^{(R)}\rangle$ and  $\langle\psi^{(L)}\vert x\vert\psi^{(L)}\rangle$ is large compared with both the $x$-variance of $\psi^{(L)}$ and that of $\psi^{(R)}$. Thus, an increase in the typical lengths of a system with a ground state conforming to (\ref{I.05}) would make this ground state even more sensitive to perturbations in the potential.

Granted, usual perturbation theory seems to disprove our main result, stated above. But this only because in this standard procedure first order perturbations in the potential are postulated and then first or higher order perturbations in the energy levels and stationary solutions are deduced. Indeed, according to D. Bohm usual perturbation theory rests on the assumption that a small change in the potential can only bring a small change in the solutions \cite{Bohm}.  An assumption we claim is not generally true, giving our results as a counterexample.

The rest of the Letter is organized as follows: In section 2 we lay out our main result and in section 3 we illustrated it with examples. Then, some plausible consequences of this result are discussed in section 4. Finally, in section 5 we lay out tentative conclusions. 
%
%
%
%
%
%
\section{An exact result for analytic potentials}
In the following pages we will be mainly concerned with an equation of the type:
 \begin{equation}\label{GRO}
           -\frac{d^{2}\psi}{dx^{2}}(x)+V(x)\psi(x)=0 
           \end{equation}
for which
   \begin{enumerate}[(i)]
 \item the \emph{potential} $V: \mathbb{R}\rightarrow\mathbb{R}$ is analytic in $\mathbb{R}$ and
   \item the eigensolution $\psi: \mathbb{R}\rightarrow\mathbb{R}$ is positive-definite, i e. 
   \begin{equation}\label{PD}
  \psi(x)>0  \quad  \forall x \in \mathbb{R} ,
   \end{equation}
and square-integrable, i. e.
  \begin{equation}\label{SQI}
   0<\int_{-\infty}^{\infty}\psi_{E}^{2}(x)dx<\infty
   \end{equation}
      \end{enumerate}
%
%
%
%
%
%
%
%
%
%
%
%
%
%
%
%
%
%
%
%
%
%
%
%
%
%
%
\begin{rmk}
\textbf{Definition.} $\mathcal{V}$ shall designate for us the set of all functions $V:\mathbb{R}\rightarrow\mathbb{R}$ complying with equation (\ref{GRO}) and conditions (I) and (II) stated above.
\end{rmk} 
%
%
%
%
%
%
%
%
%
%
%
%
%
%
%
%
%
%
%
%
%
%
%
%
%
%
%
\begin{rmk}
\textbf{Definition. }In the following we will designate by $\mathcal{B}$ the set of all analytic functions $\psi:\mathbb{R}\rightarrow\mathbb{R}$ which are positive-definite and square-integrable.
\end{rmk}      
%
%
%
%
%
%
%
%
%
%
%
%
%
%
%
%
%
%
%
%
%
%
%
%
%
%
%
%
\begin{rmk}
\textbf{Remark. }Equation (\ref{GRO}) has the form of a one-dimensional Schr\"{o}dinger stationary equation for the energy eigenvalue $E=0$. Condition (\ref{PD}) implies that $\psi$ is \emph{nodeless} and is thus the ground eigensolution of potential $V$. Consequently, the $V$ of equation (\ref{GRO}) has ground eigenenergy $E_{0}=0$. In so many words: $\mathcal{V}$ is (save for multiplicative factors of the type $m/\hbar^{2}$) the set of all analytic potentials $V:\mathbb{R}\rightarrow\mathbb{R}$ with ground energy eigenvalue $E_{0}=0$.

Moreover, for each analytic $U:\mathbb{R}\rightarrow\mathbb{R}$ with a ground eigenvalue $E_{0}$ there is a function $V=(U-E_{0})\in\mathcal{V}$ with the same eigensolutions as $U$. Thus, any potential analytic on the line will have a ground-level  square-integrable eigensolution $\psi\in\mathcal{B}$ and a properly normalized ground-level eigensolution $\Psi\in\mathcal{B}$ ( that is a $\Psi$ complying with
 \begin{equation}\label{PN00}
    \int_{-\infty}^{\infty}\Psi^{2}(x)dx=1  \quad\quad\Bigg)
    \end{equation}
  if it admits bounded states at all. 
\end{rmk}
%
%
%
%
%
%
%
%
%
%
%
%
%
%
%
%
%
%
%
%
%
%
%
%
%
%
%
%
\begin{rmk}
 \textbf{Definition. }In the following $\mathcal{A}\subset\mathcal{B}$ will stand for the set of all properly normalized functions in $\mathcal{B}$.
\end{rmk}
%
%
%
%
%
%
%
%
%
%
%
%
%
%
%
%
%
%
%
%
%
%
%
%
%
%
%
%
\begin{lem}
There exists a surjective mapping $\mathfrak{M}:\mathcal{B}\rightarrow\mathcal{V}$  such that:
    \begin{itemize} 
           \item[I)] For every $\psi\in\mathcal{B}$ there is one and only one $V=\mathfrak{M}\psi=\frac{\psi^{\prime\prime}}{\psi}\in\mathcal{V}$ for which (\ref{GRO}) is verified (the prime standing for the $x$-derivative) and 
           \item[II)]the restriction $\mathfrak{M}\vert_{\mathcal{A}}:\mathcal{A}\rightarrow\mathcal{V}$ is invertible.
           \end{itemize}
\end{lem}
%
%
%
%
%
%
%
%
%
%
%
%
%
%
%
%
%
%
%
%
%
%
%
%
%
%
%
%
\begin{pf}
As every $\psi\in\mathcal{B}$ is analytic and positive-definite in $\mathbb{R}$, then
\begin{equation}\label{MI01}
[\mathfrak{M}\psi](x)=\frac{\psi^{\prime\prime}(x)}{\psi (x)}
\end{equation}
is a well defined  expression for all  $x\in\mathbb{R}$, and by making $V(x)=\psi^{\prime\prime}(x)/\psi (x)$ equation (\ref{GRO}) is automatically satisfied. Thus, the mapping exists. We know that it is surjective because, by definition, $V$ belongs in $\mathcal{V}$  only if there exists a $\psi\in\mathcal{B}$ that complies with (\ref{GRO}). Elementary theorems warrant that  given a $V\in\mathcal{V}$ there is one and only one $\Psi\in\mathcal{A}$ that is a solution of (\ref{GRO}). Thus, the restriction $\mathfrak{M}\vert_{A}$ is invertible.
\end{pf}
%
%
%
%
%
%
%
%
%
%
%
%
%
%
%
%
%
%
%
%
%
%
%
%
%
%
%
%
\begin{lem}
 $\mathcal{B}$ is convex, that is: given $\psi , \phi\in\mathcal{B}$ then $\alpha\psi +(1-\alpha)\phi\in\mathcal{B}$ for all $\alpha\in [0,1]$.
\end{lem}
%
%
%
%
%
%
%
%
%
%
%
%
%
%
%
%
%
%
%
%
%
%
%
%
%
%
%
\begin{pf}
Any finite linear combination of analytic functions is analytic and any finite linear combination of square-integrable functions is square-integrable. Also, a finite linear combination of positive definite functions will be positive definite if all coefficients are non-negative and at least one of the coefficients is different from zero.
\end{pf}
%
%
%
%
%
%
%
%
%
%
%
%
%
%
%
%
%
%
%
%
%
%
%
%
%
%
%
%
\begin{lem}
Given any two functions $\psi^{(0)},\psi^{(1)}\in\mathcal{B}$ there exist a family $\{V_{\alpha}\}\vert_{\alpha\in [0,1]}\subset\mathcal{V}$ given by  
    \begin{equation}\label{FAM01}
       V_{\alpha}=\frac{(1-\alpha)V_{0}\psi^{(0)}+\alpha V_{1}\psi^{(1)}}{(1-\alpha)\psi^{(0)}+\alpha\psi^{(1)}} \quad ,
       \end{equation}
where, by definition
 \begin{equation}\label{FAM01.A}
V_{0}=[\mathfrak{M}\psi^{(0)}]\in\mathcal{V}
\end{equation}
 and
 \begin{equation}\label{FAM01.B}
V_{1}=[\mathfrak{M}\psi^{(1)}]\in\mathcal{V},
\end{equation}
with mapping $\mathfrak{M}$ as defined in (\ref{MI01}).  
 Each $V_{\alpha}$ has a ground solution $\psi^{(\alpha)}\in\mathcal{B}$ given by 
\begin{equation}\label{FAM02}
  \psi^{(\alpha)}=(1-\alpha)\psi^{(0)} +\alpha\psi^{(1)}\quad .
   \end{equation}
\end{lem} 
%
%
%
%
%
%
%
%
%
%
%
%
%
%
%
%
%
%
%
%
%
%
%
%
%
%
%
%
%
%
%
%
\begin{pf}
From Lemma 2, for each  $\alpha\in [0,1]$ the convex linear combination (\ref{FAM02}) belongs in $\mathcal{B}$. By Lemma 1  there is one and only one $V_{\alpha}\in\mathcal{V}$, namely
\begin{equation}\label{FAM03}
V_{\alpha}=\mathfrak{M}\psi^{(\alpha)},
\end{equation}
with mapping $\mathfrak{M}$ as defined  in (\ref{MI01}),  for which  $\psi^{(\alpha)}$ is a ground eigensolution.  Substituting (\ref{FAM02}) in (\ref{FAM03}) gives:
 \begin{equation}\label{FAM04}
       V_{\alpha}=\frac{(1-\alpha)\psi^{(0)\prime\prime}+\alpha\psi^{(1)\prime\prime}}{(1-\alpha)\psi^{(0)}+\alpha\psi^{(1)}} \quad ,
       \end{equation}
plugging definitions (\ref{FAM01.A}) and (\ref{FAM01.B}) in (\ref{FAM04}) gives us, finally, equation (\ref{FAM01}).
\end{pf}
%
%
%
%
%
%
\begin{lem}
For each $V\in\mathcal{V}$ there exists an uncountable set of families $\{ V_{\alpha}\}\vert_{\alpha\in(0,1)}\subset\mathcal{V}$ of the form (\ref{FAM01}) to which it belongs.
\end{lem}
%
%
%
%
\begin{pf}
From Lemma 1 we have that each given $V\in\mathcal{V}$ has a unique properly normalized positive-definite ground level eigensolution
\begin{equation}\label{LCEF}
\Phi=(\mathfrak{M}\vert_{\mathcal{A}})^{-1}V\quad .
\end{equation}
Now, take any function $f\in\mathcal{B}$ such that
\begin{equation}\label{nwst01}
0< f(x)< 1 \quad \forall x\in\mathbb{R}
\end{equation} 
and define, for the $\Phi$ of equation (\ref{LCEF}), the functions
\begin{equation}\label{nwst02}
\phi^{(0)}(x)=2f(x)\Phi(x)\forall x\in\mathbb{R}
\end{equation}
and 
\begin{equation}\label{nwst03}
\phi^{(1)}(x)=2(1-f(x))\Phi(x)\forall x\in\mathbb{R}
\end{equation}
so that the convex linear combination
\begin{equation}\label{nwst04}
\phi^{(\alpha=1/2)}=\frac{1}{2}\phi^{(0)}+\frac{1}{2}\phi^{(1)}=\Phi .
\end{equation}
From Schwartz inequality and equations (\ref{nwst01}), (\ref{nwst02}) and (\ref{nwst03}) it follows that $\phi^{(0)}$ and $\phi^{(1)}$ both belong in $\mathcal{B}$. Thus, according to Lemma 3 there exists a family of potentials of the form (\ref{FAM01}) with
\begin{equation}\label{nwst05}
V_{0}=\mathfrak{M}\phi^{(0)}\quad \textrm{ , }\quad V_{1}=\mathfrak{M}\phi^{(1)}\quad\textrm{ and }\quad V=V_{\alpha=1/2}=\mathfrak{M}\Phi\quad\textrm{ . } 
\end{equation}
\end{pf}
%
%
%
%
%
%
%
%
%
%
%
%
%
%
%
%
%
%
%
%
%
%
%
%
%
%
%
%
\begin{lem}
For any given $M>0$ there exists a family of potentials $\{V_{M,\alpha}\}_{\alpha\in(0,1)}\subset\mathcal{V}$ with ground eigensolutions $\psi^{(M,\alpha)}\in\mathcal{B}$ and a pair of real numbers $\alpha_{M},\beta_{M}\in(0,1)$ that comply with 
\begin{equation}\label{II.N.0}
\frac{Q[\psi^{(M,\alpha_{M})}]}{Q[\psi^{(M,\beta_{M})}]}>M
\end{equation}
where quotient $Q$ is as defined in (\ref{I.03}).
\end{lem}
%
%
%
%
%
%
%
%
%
%
%
%
%
%
%
%
%
%
%
%
%
%
%
%
\begin{pf}
Take any two functions $\phi^{(0)},\phi^{(1)}\in\mathcal{A}$ and define, for any given $X>0$, the family $\{\phi^{(X,\alpha)}\}_{\alpha\in(0,1)}\subset\mathcal{B}$ by
\begin{equation}\label{II.N.1}
\phi^{(X,\alpha)}(x)=\alpha \phi^{(1)}(x-X)+(1-\alpha)\phi^{(0)}(x+X)\quad .
\end{equation}

Now, by hypothesis both $\phi^{(0)}$ and $\phi^{(1)}$ belong in $\mathcal{A}$ so that both are square-integrable, and thus we can find for any $\delta>0$ a $X_\delta>0$ such that 
\begin{equation}\label{II.N.7}
 \int_{0}^{\infty}[\phi^{(0)}(x+X_{\delta})]^{2}dx<\delta^{2}
\end{equation}
and
\begin{equation}\label{II.N.8}
 \int_{-\infty}^{0}[\phi^{(1)}(x-X_{\delta})]^{2}dx<\delta^{2}\quad .
\end{equation}
Furthermore, from (\ref{II.N.7}), (\ref{II.N.8}) and the Schwartz inequality we have that
 \begin{equation}\label{II.N.9}
 \int_{0}^{\infty}\phi^{(0)}(x+X_{\delta})\phi^{(1)}(x-X_{\delta})dx<\delta
\end{equation}
and
 \begin{equation}\label{II.N.10}
 \int_{-\infty}^{0}\phi^{(0)}(x+X_{\delta})\phi^{(1)}(x-X_{\delta})dx<\delta\quad .
\end{equation}
Thus, for any given $\delta>0$ there exists an $X_{\delta}>0$ such that 
 \begin{equation}\label{II.N.11}
\frac{\alpha^{2}+2\delta+\delta^{2}}{(1-\delta^{2})(1-\alpha)^{2}}> Q[\Phi^{(\delta,\alpha)}]>\frac{(1-\delta^{2})\alpha^{2}}{(1-\alpha)^{2}+2\delta+\delta^{2}}
\end{equation}
and
 \begin{equation}\label{II.N.12}
\frac{(1-\alpha)^{2}+2\delta+\delta^{2}}{(1-\delta^{2})\alpha^{2}}> Q[\Phi^{(\delta,1-\alpha)}]>\frac{(1-\delta^{2})(1-\alpha)^{2}}{\alpha^{2}+2\delta+\delta^{2}}
\end{equation}
for any $\alpha\in(0,1)$, where, by definition
 \begin{equation}\label{II.N.13}
\Phi^{(\delta,\alpha)}(x)=\phi^{(X_{\delta},\alpha)}(x)\quad .
\end{equation}
We thus have for each $\delta\in(0,1)$ a family of functions $\{\Phi^{(\delta,\alpha)}\}_{\alpha\in(0,1)}\subset\mathcal{B}$ such that 
\begin{equation}\label{II.N.14}
\frac{Q[\Phi^{(\delta,\alpha)}]}{Q[\Phi^{(\delta,1-\alpha)}]}>\Bigg(\frac{(1-\delta^{2})\alpha^{2}}{(1-\alpha)^{2}+2\delta+\delta^{2}}\Bigg)^{2}
\end{equation}
Now, for any given $M>0$ define the positive real number 
 \begin{equation}\label{II.N.15}
\delta_{M}=\frac{a_{M}b_{M}}{1+a_{M}+b_{M}}
\end{equation} 
where $a_{M},b_{M}$ are shorthand for
 \begin{equation}\label{II.N.16}
a_{M}=\frac{M}{M+4}\Bigg(-1+\sqrt{(\frac{M}{M+4})^{2}+1}\Bigg) 
\end{equation}
and  \begin{equation}\label{II.N.17}
b_{M}=\frac{M+1}{M+2}\Bigg(-1+\sqrt{\frac{(M+2)M}{(M+1)^{3}}+1}\Bigg)
\end{equation}
Equations (\ref{II.N.15}), (\ref{II.N.16}) and  (\ref{II.N.17}) warrant, first, that $1>\delta_{M}>0$. In second place, this equations warrant that the $\alpha_{M}$ defined by the expression 
\begin{equation}\label{II.N.17.A}
\alpha_{M}=\frac{M+1-\sqrt{(M+1)^{2}-(M+1)(M+\delta_{M}^{2})(1+\delta_{M})^{2}}}{M+\delta_{M}^{2}}
\end{equation}
is a positive real number, one of the solutions to the equation
\begin{equation}\label{II.N.18}
\frac{(1-\delta_{M}^{2})\alpha^{2}}{(1-\alpha)^{2}+\delta_{M}^{2}+2\delta_{M}}=M+1\quad \textrm{ . }
\end{equation}
Also, equations (\ref{II.N.15}), (\ref{II.N.16}) and  (\ref{II.N.17}) necessarily imply that $0<\alpha_{M}<1$.

For each $M>0$ we define the family of functions  $\{\psi^{(M,\alpha)}\}_{\alpha\in(0,1)}\subset\mathcal{B}$ as
 \begin{equation}\label{II.N.19}
\psi^{(M,\alpha)}(x)\quad=\Phi^{(\delta_{M},\alpha)}(x)
\end{equation}
so that, according with (\ref{II.N.14}), 
\begin{equation}\label{II.N.20}
\frac{Q[\psi^{(M,\alpha)}]}{Q[\psi^{(M,1-\alpha)}]}>\Bigg(\frac{(1-\delta_{M}^{2})\alpha^{2}}{(1-\alpha)^{2}+2\delta_{M}+\delta_{M}^{2}}\Bigg)^{2}
\end{equation}
for all $\alpha\in(0,1)$ and 
\begin{equation}\label{II.N.21}
\frac{Q[\psi^{(M,\alpha_{M})}]}{Q[\psi^{(M,1-\alpha_{M})}]}>(M+1)^{2}>M .
\end{equation}
Finally, with the use of the mapping defined in Lemma 1  we now define, for each $M>0$ and each $\alpha\in(0,1)$, the potential $V_{M,\alpha}=\mathfrak{M}\psi^{(M,\alpha)}$
\end{pf}
%
%
%
%
%
%
%
%
%
%
%
%
%
%
%
%
%
%
\begin{lem}
For each $\Psi\in\mathcal{A}$ (with corresponding potential  $V\in\mathcal{V}$) and each $k>0$ there exist a $\Psi^{[k]}\in\mathcal{A}$, defined by 
\begin{equation}\label{R1}
\Psi^{[k]}(x)=\sqrt{k}\Psi(kx)\quad\forall x\in\mathbb{R}
\end{equation}
which is a solution of the equation 
\begin{equation}\label{R101}
-\Psi^{[k]\prime\prime}+V_{[k]}\Psi^{[k]}=0
\end{equation}
where $V_{[k]}\in\mathcal{V}$ is given by
\begin{equation}\label{R102}
V_{[k]}(x)=k^{2}V(kx)\quad\forall x\in\mathbb{R}\quad .
\end{equation}
\end{lem}
%
%
%
%
%
%
%
%
%
%
%
%
%
%
%
%
%
%
%
%
%
%
%
\begin{pf}
As the product of two analytic functions is an analytic function, and the composition of any two analytic functions is an analytic function, then the $\Psi^{[k]}$ defined in (\ref{R1}) is analytic whenever $\Psi$ belongs in $\mathcal{A}$. Also, as $\sqrt{k}>0$, then $\Psi^{[k]}$ is positive-definite if $\Psi$ belongs in $\mathcal{A}$. Moreover, as any $\Psi\in\mathcal{A}$ is normalized we have that:
\begin{equation}\label{R103}
\int_{-\infty}^{\infty} \Psi^{[k]2}(x) dx =\int_{-\infty}^{\infty} k\Psi^{2}(kx) dx= 1
\end{equation}
for any $k>0$. Thus, $\Psi^{[k]}$ belongs in $\mathcal{A}$ whenever  $\Psi$ belongs in $\mathcal{A}$.

The chain rule gives, when applied to the second derivative of (\ref{R1}), the result:
\begin{equation}\label{R104}
\Psi^{[k]\prime\prime}(x)=k^{2}\sqrt{k}\Psi^{\prime\prime}(kx)
\end{equation}
Dividing (\ref{R104}) by (\ref{R1}) gives
\begin{equation}\label{R105}
\frac{\Psi^{[k]\prime\prime}(x)}{\Psi^{[k]}(x)}=\frac{k^{2}\Psi^{\prime\prime}(kx)}{\Psi(kx)}=k^{2}V(kx)\ .
\end{equation}
\end{pf}
%
%
%
%
%
%
%
%
%
%
%
%
%
%
%
%
%
%
%
%
\begin{lem}
The quotient $Q$, defined in equation (\ref{I.04}) is invariant under the family of mappings $\Psi\rightarrow\Psi^{[k]}$ defined in Lemma 6, that is 
\begin{equation}\label{I002}
Q[\Psi^{[k]}]=Q[\Psi] \quad\forall \Psi\in\mathcal{A}\quad , \quad\forall k>0
\end{equation}
\end{lem}
%
%
%
%
%
%
%
%
%
%
%
%
%
%
%
%
%
%
%
%
\begin{pf}
For any two real numbers $a,b\in\mathbb{R}$ we have that, according with definition 
\begin{equation}\label{I003}
\int_{a}^{b}\Big(\Psi^{[k]}(x)\Big)^{2}dx=\int_{ka}^{kb}\Big(\Psi(y)\Big)^{2}dy
\end{equation} 
so that, for any $a\in\mathbb{R}$
\begin{equation}\label{I003A}
\int_{a}^{\infty}\Big(\Psi^{[k]}(x)\Big)^{2}dx=\int_{ka}^{\infty}\Big(\Psi(x)\Big)^{2}dx
\end{equation} 
and
\begin{equation}\label{I004}
\int_{-\infty}^{a}\Big(\Psi^{[k]}(x)\Big)^{2}dx=\int_{-\infty}^{ka}\Big(\Psi(x)\Big)^{2}dx
\end{equation} 
Thus 
\begin{equation}\label{I004A}
\int_{0}^{\infty}\Big(\Psi^{[k]}(x)\Big)^{2}dx=\int_{0}^{\infty}\Big(\Psi(x)\Big)^{2}dx\end{equation}
and 
\begin{equation}\label{I004B}
\int_{-\infty}^{0}\Big(\Psi^{[k]}(x)\Big)^{2}dx=\int_{-\infty}^{0}\Big(\Psi(x)\Big)^{2}dx
\end{equation}
Equation (\ref{I002}) follows from (\ref{I004A}) and (\ref{I004B}).
\end{pf}
%
%
%
%
%
%
%
%
%
%
%
%
%
%
%
\begin{thm}
For any given $\epsilon>0$ and $M>0$ there exists a family of potentials $\{V_{\epsilon,M,\alpha}\}\vert_{\alpha\in(0,1)}\subset\mathcal{V}$ with ground eigensolutions $\Psi^{(\epsilon,M,\alpha)}\in\mathcal{B}$ and  such that 
\begin{equation}\label{II.ZZ.01}
\vert V_{\epsilon,M,\alpha}(x)-V_{\epsilon,M,\beta}(x)\vert<\epsilon \quad \forall x\in\mathbb{R}
\end{equation}
for all $\alpha,\beta\in(0,1)$. Furthermore, there exists $\alpha_{M},\beta_{M}\in(0,1)$ such that
\begin{equation}\label{II.ZZ.03}
\frac{Q[\Psi^{(\epsilon,M,\alpha_{M})}]}{Q[\Psi^{(\epsilon,M,\beta_{M})}]}>M
\end{equation}
with $Q$ as defined in (\ref{I.04}).
\end{thm}
%
%
%
%
%
%
%
%
%
%
%
\begin{pf}
Choose any $\phi^{(0)},\phi^{(1)}\in\mathcal{A}$ with bounded potentials $W_{0}=\mathcal{M}\phi^{(0)}\in\mathcal{V}$ and $W_{1}=\mathcal{M}\phi^{(1)}\in\mathcal{V}$. Then, for each $X>0$ the potentials
\begin{equation}\label{II.ZZ.04.1}
W_{X,1}(x)=W_{1}(x-X) \textrm{ and }W_{X,0}(x)=W_{0}(x+X) 
\end{equation}
are also bounded so that there exists, for each $X>0$, a $B_{X}>0$ such that
\begin{equation}\label{II.ZZ.04.2}
\vert W_{X,1}(x)-W_{X,0}(x)\vert<B_{X} \quad \forall x\in\mathbb{R}\quad .
\end{equation}
Then
\begin{equation}\label{II.ZZZ.05}
\vert W_{X,\alpha}(x)-W_{X,\beta}(x)\vert<B_{X} \quad \forall x\in\mathbb{R}\quad \forall\alpha,\beta\in(0,1) 
\end{equation}
with $W_{X,\alpha}$ defined, for each $X>0$ and each $\alpha\in(0,1)$, by
\begin{equation}
W_{X,\alpha}=\mathfrak{M}\{\alpha\phi^{(1)}(x-X)+[1-\alpha]\phi^{(0)}(x+X)\} \ .
\end{equation}
Consequence of Lemma 5 and equation (\ref{II.ZZZ.05}) we can affirm that for each $M>0$ there exists a family of bounded functions  $\{U_{M,\alpha}\}\vert_{\alpha\in(0,1)}\subset\mathcal{V}$ with ground eigensolutions $\psi^{(M,\alpha)}\in\mathcal{A}$ and a pair of numbers numbers $\alpha_{M},\beta_{M}$ that satisfy 
(\ref{II.N.0}), and a bound $C_{M}>0$ such that
\begin{equation}\label{II.ZZZ.06}
\vert U_{M,\alpha}(x)-U_{M,\beta}(x)\vert<C_{M} \quad \forall x\in\mathbb{R}\quad .
\end{equation}
For each $\epsilon>0$ and each $M>0$ now define the number
\begin{equation}\label{II.ZZZ.07}
K_{M,\epsilon}=\sqrt{\frac{\epsilon}{C_{M}}}
\end{equation}
and then define the ground eigenfunctions
\begin{equation}\label{II.ZZZ.08}
\Psi^{(\epsilon,M,\alpha)}(x)=\psi^{[K_{\epsilon,M}](M,\alpha)}(x)=\sqrt{K_{\epsilon,M}}\phi^{(M,\alpha)}(K_{\epsilon,M}\ x)
\end{equation}
with their corresponding potentials $V_{\epsilon,M,\alpha}=\mathfrak{M}\Psi^{(\epsilon,M,\alpha)}$. From (\ref{R102}) and (\ref{II.ZZZ.06}) we then have that
\begin{equation}\label{II.ZZZ.09}
\vert V_{M,\epsilon,\alpha}(x)-V_{M,\epsilon,\beta}(x)\vert<\epsilon\quad \forall x\in\mathbb{R}\quad \forall \alpha,\beta\in(0,1).
\end{equation}
As the $\psi^{(M,\alpha)}$ comply with (\ref{II.N.0}), then Lemma 7 tells us that the $\Psi$ defined in (\ref{II.ZZZ.08}) comply with (\ref{II.ZZ.03}) for the same values of $\alpha_{M}$ and $\beta_{M}$.
\end{pf} 
%
%
%
%
%
%
%
%
%
%
%
%
%
%
%
%
%
%
%
%
%
%
%
%
%
%
%
%
%
%
\section{Examples}
\begin{figure}[h!t]
\begin{center}
\includegraphics[width=0.5\textwidth]{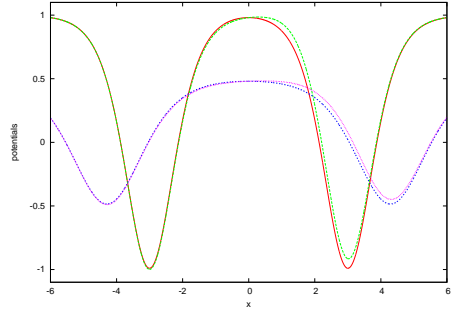}
\end{center}
\caption{Examples of the potentials of equation (\ref{EX.05}). In all cases $D=3.00$. The red ($\alpha=0.5$) and green ($\alpha=0.1$) curves correspond to potentials with $k=1.00$ while blue ($\alpha=0.5$) and magenta ($\alpha=0.1$) curves are cases in which $k=0.7$}
\end{figure}
Starting from the ground eigenfunction
\begin{equation}\label{EX.01}
\psi(x)=\textrm{ sech }(x)
\end{equation}
of the one-soliton potential 
\begin{equation}\label{EX.02}
U(x)=-2\textrm{sech}^{2} (x)\ +1 
\end{equation}
with ground energy level $E_{0}=0$, let us define for each $k>0$, $D>0$ and $\alpha\in(0,1)$ the function
\begin{equation}\label{EX.03}
\mathcal{B}\ni\psi^{(k,D,\alpha)}(x)=\alpha\textrm{ sech }(kx-D)+(1-\alpha)\textrm{ sech }(kx+D)
\end{equation}
a ground state eigensolution to the potential
\begin{equation}\label{EX.04} 
V_{k,D,\alpha}(x)=[\mathfrak{M}\psi^{(k,D,\alpha)}](x) ,
\end{equation}
which can calculated from (\ref{MI01}) and (\ref{EX.03}), obtaining the result
\begin{equation}\label{EX.05}
V_{k,D,\alpha}(x)=k^{2}-2k^{2}\frac{\alpha\textrm{ sech}^{3}(kx-D)+(1-\alpha)\textrm{ sech}^{3}(kx+D)}{\alpha\textrm{ sech} (kx-D)+(1-\alpha)\textrm{ sech} (kx+D)} .
\end{equation}
\begin{figure}[h!t]
\begin{center}
\includegraphics[width=0.5\textwidth]{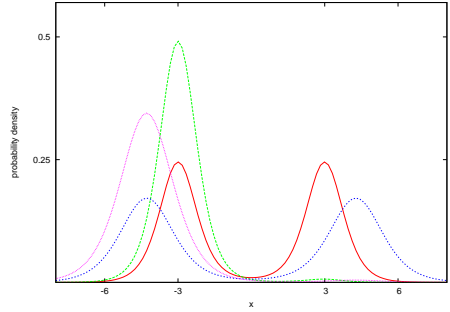}
\caption{The probability densities of the ground states corresponding to the  potentials depicted in figure 1. In all cases $D=3$. The red ($\alpha=0.5$) and green  ($\alpha=0.1$) curves corresponds to the value $k=1.00$ while blue ($\alpha=0.5$) and magenta ($\alpha=0.1$) represent the probability densities for the case $k=0.70$}
\end{center}
\end{figure}
From the known properties of the hyperbolic secant and definition (\ref{EX.05}) we have that for any $k>0$, any $D>0$ and any $\alpha,\beta\in(0,1)$ 
\begin{equation}\label{EX.06}
\vert V_{k,D,\alpha}(x)- V_{k,D,\beta}(x)\vert<4k^{2}
\end{equation}
so that given any $\epsilon>0$ there exist $k_{\epsilon}=(\epsilon/4)^{1/2}>0$ such that 
\begin{equation}\label{EX.07}
\vert V_{k_{\epsilon},D,\alpha}(x)- V_{k_{\epsilon},D,\beta}(x)\vert<\epsilon
\end{equation}
The properties of the hyperbolic secant also provide us with an exact expression for the quotient $Q$ for this functions:
\begin{equation}\label{EX.08}
Q[\psi^{(k,D,\alpha)}]=\frac{\alpha^{2}e^{D}+(1-\alpha)^{2}e^{-D}+2\alpha(1-\alpha)D/\sinh D}{\alpha^{2}e^{-D}+(1-\alpha)^{2}e^{D}+2\alpha(1-\alpha)D/\sinh D}
\end{equation}
so that
\begin{equation}\label{EX.09}
\lim_{\alpha\rightarrow 1}\frac{Q[\psi^{(k,D,\alpha)}]}{Q[\psi^{(k,D,1-\alpha)}]}=e^{4D}\quad .
\end{equation}
Thus, by defining for each $M>0$
\begin{equation}\label{EX.10}
D_{M}=\ln\Bigg(M+1\Bigg)^{1/4}
\end{equation}
we guarantee the existence of an $\alpha_{M}\in(0,1)$ such that
\begin{equation}\label{EX.11}
\frac{Q[\psi^{(k,D_{M},\alpha_{M})}]}{Q[\psi^{(k,D_{M},1-\alpha_{M})}]}>M
\end{equation}
for all $k>0$. 

Figures 1 and 2 illustrate the sensitivity of ground eigenfunctions (\ref{EX.03}) to small perturbations in  potentials (\ref{EX.05}).
 
This first example was chosen so that the parity symmetry of an \emph{unpertubed} potential (the case for $\alpha=1/2$) was broken by small asymmetric perturbations. Our next and last example (which we will not work out in full) shows how the sensitivity of the ground eigensolution can appear even when parity is absent from the unperturbed case. Indeed, just consider the convex linear combinations of the form
\begin{equation}\label{EX.12}
\phi^{k,D,\alpha}(x)=\alpha \textrm{ sech }(kx+D)+(1-\alpha)\nu(kx-D)
\end{equation}
where
\begin{equation}\label{EX.13}
\nu(x)=e^{x/2}\textrm{ sech } (x)
\end{equation} 
is a  ground eigensolution of the hyperbolic Rosen-Morse potential:   
\begin{equation}\label{EX.14}
V_{\textrm{\tt{ RM }}}(x)=\frac{5}{4}-2\textrm{ sech }^{2} x+\tanh x ,
\end{equation}
as can be checked by direct substitution.

Clearly, no combination of values of $k>0$, $D>0$ and $\alpha\in(0,1)$ can make ground eigensolution (\ref{EX.12}) (and thus its potential) symmetric. Yet, by now it should be clear that the behaviour of the potentials and ground eigenstates is very similar to the one in the previous example: in both cases each potential in the family consists of two wells separated by a certain distance. The sensitivity of the ground eigensolutions depends not on the exact details of each potential but rather on the separation between the wells. For bigger distances, each small change in the relative strength between the wells entails an inordinate change in the probability of finding the particle in one or the other well.
      
We note in passing that other, more toilsome examples can be worked out starting from different convex linear combinations of hyperbolic Scarf, hyperbolic Rosen-Morse and other exactly solvable bounded potentials. Indeed, the sensitivity of the ground eigenfunction can be shown to exist even without any knowledge of the excited spectra, just by judiciously choosing functions $\psi^{(0)},\psi^{(1)}\in\mathcal{B}$ with bounded $\mathfrak{M}\psi$ potentials.
\section{Discussion}
Lemma 4 shows us that there is nothing inherently pathological about the type of potentials under consideration in his Letter: every ground eigensolution can be written as the convex linear combination of two different square-integrable functions (even if this latter functions bear, in most instances, no physical meaning at all) and every potential with (an at least) partially bounded spectrum can be seen to be part of one of our $\{V_{\alpha}\}\vert_{\alpha\in(0,1)}$ families. We thus consider that the arguments laid out in the previous pages show in a cogent manner that non-relativistic quantum theory does not disavow the possibility of systems endowed with ground eigensolutions that are extremely sensitive to small perturbations in the potential. Moreover, the analyticity of the potential, and the very stern condition (\ref{I.02}) can be relaxed in order to obtain more general versions of our main result. Indeed, the only truly indispensable condition for the extreme sensitivity of the grounds eigensolution seems to be the existence of two wells separated by a distance that is in some sense large.  
  
Consequence of the above, we believe it is reasonable to consider that a version of the effect just described may be at play whenever an experiment shows the existence of a large, unexplained asymmetry in a distribution probability. Thus, for example, when considering the self-trapping phenomena occurring  Bose-Einstein condensates \cite{Smerzietal1997,Milburnetal1997,Albiezetal2005}, it may be worthwhile to investigate if the (admittedly non-linear) Gross-Pitaevskii equation governing this systems admits non-standard perturbations similar to the ones expounded in the previous pages. If this were to be true, then  uncontrollable small perturbations in the confining potential may enhance the self-trapping phenomenon. In a similar vein, we conjecture that small asymmetries in the composition profile may play a role in the observed accumulation of localized electrons in quantum well heterostructures composed of narrow-gap semiconductors\cite{Ishida2010}. We consider that discussing other possible implications of the results just expounded would be both premature and highly speculative at this stage.  

Finally, we have to mention two factors that could hinder the experimental detection of the extreme sensitivity we have been alluding to. The first of this is the mingling of the ground and first excited state into an effective two level system. Indeed, in the type of double-well potentials under discussion, the first excited eigensolution, $\psi_{E_{1}}$, can always be approximated \cite{Munozetal2012, Munozetal2013} by
\begin{equation}
\psi_{E_{1}}\approx n_{L}\psi^{(L)}-n_{R}\psi^{(R)}
\end{equation}
and as their overlap diminishes, the $\psi^{(L)}$ and $\psi^{(R)}$ become the \emph{increasingly independent} (so to speak) solutions of a low-laying two-level system. 

The second hindrance we have mentioned stems from the scaling of the potentials  given by equation (\ref{R102}), which indicates that sensitivity is favored by \emph{slow} particles, i. e. particles with small expected values of kinetic energy.

\section{Conclusions.}
In the preceding pages we have shown that the non-relativistic quantum theory of structureless particles moving on the line admits the existence of stationary perturbations differing from the standard type. Our models predict the existence of dramatic changes in the ground probability density with only minor perturbations in the potential, when this potential is bounded and consists in two wells separated by a distance large enough as to make the overlap between the independent well solutions negligible. This is a falsifiable prediction  stemming only from text-book non-relativistic quantum theory that may have implications for condensed matter physics and possibly other fields.  
\section{Acknowledgements}
The support of SNI-CONACYT is duly acknowledged. 

\end{document}